\documentclass[twocolumn,showpacs,showkeys,superscriptaddress]{revtex4}
\usepackage{graphicx}
\usepackage{amsmath}
\usepackage{amsfonts}
\usepackage{amssymb}
\usepackage{mathrsfs}

\def\openone{\leavevmode\hbox{\small1\kern-3.8pt\normalsize1}}
\def\N{\leavevmode\hbox{ Z \kern-8 pt\normalsize{Z}}}
\def\openone{\leavevmode\hbox{\small1\kern-3.8pt\normalsize1}}
\def\openJ{\leavevmode\hbox{J \kern-9.5pt\normalsize J}}
\def\openS{\leavevmode\hbox{ S \kern-9.3pt\normalsize S}}
\newcommand{\bb}{\begin{equation}}
\newcommand{\ee}{\end{equation}}
\newcommand{\eqb}{\begin{eqnarray}}
\newcommand{\eqf}{\end{eqnarray}}

\usepackage{color}

\begin{document}

\title{New non-linear modified massless Klein--Gordon equation}

\author{Felipe A. Asenjo}
\email{felipe.asenjo@uai.cl}
\affiliation{UAI Physics Center, Universidad Adolfo Ib\'a\~nez, Santiago, Chile.}
\affiliation{Facultad de Ingenier\'{\i}a y Ciencias,
Universidad Adolfo Ib\'a\~nez, Santiago, Chile.}
\author{Sergio A. Hojman}
\email{sergio.hojman@uai.cl}
\affiliation{UAI Physics Center, Universidad Adolfo Ib\'a\~nez, Santiago, Chile.}
\affiliation{Departamento de Ciencias, Facultad de Artes Liberales,
Universidad Adolfo Ib\'a\~nez, Santiago, Chile.}
\affiliation{Departamento de F\'{\i}sica, Facultad de Ciencias, Universidad de Chile,
Santiago, Chile.}
\affiliation{Centro de Recursos Educativos Avanzados,
CREA, Santiago, Chile.}

\begin{abstract}
The massless Klein--Gordon equation on arbitrary curved backgrounds allows for solutions which develop ``tails'' inside the light cone and, therefore, do not strictly follow null geodesics as discovered by DeWitt and Brehme almost sixty years ago. A modification of the massless Klein--Gordon equation is presented which always exhibits null geodesic propagation of waves on arbitrary curved spacetimes. This new equation is derived from a Lagrangian which exhibits current--current interaction. Its non--linearity is due to a self--coupling term which is related to the quantum mechanical Bohm potential.
\end{abstract}


\maketitle

\section{Introduction}

It is well-known that point structureless particles which move on arbitrary background gravitational fields follow geodesics according to the Equivalence Principle (EP) of General Relativity. In particular, massless particles move along null geodesics. For example, the crucial light deflection observational results were predicted and matched by considering photons as massless particles moving along null geodesics. However, these results are valid under the assumption of  weak and smooth gravitational fields and light behaving in the geometrical optics limit (where light propagation is treated as the motion of massless spinless particles).

The caveat is that point structureless particles do not exist in Nature. If the particle has structure, then its motion may be completely different from the one followed by point particles. This is the case of spinning massive particles moving on arbitrary curved backgrounds that do not, in general, follow geodesics \cite{mat,pap1,hojman1,gane1,gane2,armaza}. On the other hand, Nature is better described in terms of fields which are neither point--like nor structureless entities, in the sense that they in general correspond to extended propagating objects that in addition can carry spin. Therefore, the dynamics of spinning particles and fields differs from the one for spinless point particles, and one should  expect that the EP may not be applied in those cases. The reason is that for extended (not point--like) and/or structured  (such as  spinnning) objects, several geodesic curves cross a sufficiently extended region of the body, making it subject to tidal forces. Thus, the geodesic path does not make sense for physical objects different from point-like ones.

In the spirit of the discussion presented above, one can expect new features in the dynamics of any field coupled to gravity.
The study of the propagation of massless fields on arbitrary curved backgrounds has been the subject of research for a long time \cite{mcglen, dwb,faraoni,Yagdjian,Noonan, Noonan2, Noonan3, Noonan4,Anile,zecca,Noonan5,Noonan6, sonego, sonego2, faraoni2}.
In 1960, DeWitt and Brehme \cite{dwb} found solutions both for the Klein--Gordon and the Maxwell fields on a curved background which exhibit ``tails'' inside the light cone, meaning that the propagators for these fields do not vanish inside the light cone. These ``tails'' are a pure general relativistic effect, showing how the spreading of the wave field over spacetime can modify its dynamics. These tails can be obtained trhough the propagators of  the fields , the Green function \cite{hamadar, john}, for  linear wave equations in different backgrounds.

Recently, several new results for the propagation of (classical and quantum) electromagnetic waves on arbitrary curved backgrounds suggest non--null geodesic and polarization dependent  propagation of waves \cite{khri,hollow,Drummond,Klein,danielshore,danielshore2,danielshore3,danielshore4,Khriplovich,heylH,agullo,asenjohojmanLight,asenjohojmanLight2}. Similar results arise in the study of the propagation of spin field waves \cite{ohkuwa, velo}. Up to now, these new results seem to be neither confirmed nor refuted by experiments or observations.

One can study many of the ``tails'' effects for field dynamics  even at the simplest level. That is the case of  the Klein--Gordon equation for a massless charged field $\Phi=\Phi(x^\mu)$ on a background curved spacetime
\begin{equation}\label{KG0}
\Box \Phi=0\, ,
\end{equation}
with the operator $\Box\equiv\partial_\mu\left(\sqrt{-g} g^{\mu\nu}\partial_\nu \right)/{\sqrt{-g}}$ (for scalar fields), where $\partial_\mu$ is a partial derivative, $g_{\mu\nu}$ is the metric of an arbitrary curved background (with inverse $ g^{\mu\nu}$ and determinant $g$). Greek indices run from 0 to 3.

Instead of calculating the Green function \cite{john}
 for the field $\Phi$  satisfying \eqref{KG0}, in order to obtain the ``tails'', we can simply study an analogous procedure for  waves that show this feature. Let us assume a scalar wave solution for the field
\begin{equation}\label{complexdec}
\Phi= \phi\, e^{i S}\, ,
\end{equation}
where its (positive) absolute value is $\phi= \phi(x^\alpha)=\sqrt{\Phi^*\Phi}$, corresponding to the  amplitude of the wave. The real phase of the wave is $S=S(x^\beta)$, defining the four--wavevector $k_\mu$ as
\begin{equation}\label{kmu}
k_\mu =\partial_\mu S\, ,
\end{equation}
where $k_0$ is the wave frequency, and $\vec k=\vec \nabla S$ defines the three-dimensional wave vector.
It is the wavevector \eqref{kmu} that defines the propagation nature of this scalar wave.
If this scalar wave, describing a massless field,
would move along null geodesics, then $k_\mu k^\mu=g^{\mu\nu}k_\mu k_\nu=0$ must be a solution of Eq.~\eqref{KG0}.

Using the field \eqref{complexdec} in  Eq.~\eqref{KG0}, we find the equations
\begin{eqnarray}\label{firstsetEqs}
k_\mu k^\mu=\frac{\Box\phi}{\phi}\, ,\qquad
\nabla_\mu\left(\phi^2 k^\mu\right)=0\, ,
\end{eqnarray}
where $\nabla_\mu$ is a covariant derivative. Eqs.~\eqref{firstsetEqs} are general for any background curved spacetime. The first one determines how the waves propagates and it is usually referred as the dispersion relation. The second equation is the propagation equation for the scalar amplitude and it corresponds to the conservation of the rays of the scalar wave.

It is not difficult to check that in flat spacetime $k_\mu k^\mu=0$ is a solution of  Eqs.~\eqref{firstsetEqs} for constant amplitude $\phi$. However, in curved spacetime the solution is not trivial and it depends on the metric. Thus, in general, in curved spacetime,  Eqs.~\eqref{firstsetEqs} allows for $k_\mu k^\mu\neq 0$ for a massless field, implying its non null geodesic propagation. This is equivalent to the ``tails'' of the field.

This feature of  Eqs.~\eqref{firstsetEqs} is ackward, and several attempts to get a modified massless Klein--Gordon
that produces waves following null geodesics have been tried. One of them has been successful, at leat for  some curved backgrounds \cite{Noonan,Noonan2}.
In it, the massless Klein-Gordon equation on a curved background is obtained by modifying the minimal coupling of matter to gravity in a conformal manner. However, this procedure is known to produce scalar waves moving along null geodesics only for specific metric backgrounds \cite{Noonan}. We sketch this result in the next section.

One can also wonder if there is possible a modification to the massless Klein--Gordon equation such that every scalar wave follows null geodesics in any curved background metric, such that $k_\mu k^\mu=0$ always. It is the aim of this work to show that such modification does indeed exist. The new Klein--Gordon equation is a non--linear modification which is conformally invariant.
Thereby, the propagation of charged massless spinless (modified) Klein--Gordon fields follow null geodesics on any curved spacetime.
We study such equation in Sec.~\ref{secSup}

\section{Conformally invariant massless Klein--Gordon equation}

The conformally invariant massless Klein--Gordon equation can avoid the ``tails'' for some background metrics. This massless Klein--Gordon equation on a curved background is written by modifying the minimal coupling prescription in order to get a conformally invariant equation \cite{wald}
\begin{equation}\label{KG1}
\Box \Phi-\frac{R}{6}\Phi=0\, ,
\end{equation}
where $R$ is a Ricci scalar of curvature. For scalar waves \eqref{complexdec}, the previous Eq.~\eqref{KG1} can be cast as
\begin{eqnarray}\label{firstsetEqs2}
k_\mu k^\mu=\frac{\Box\phi}{\phi}-\frac{R}{6}\, ,\qquad
\nabla_\mu\left(\phi^2 k^\mu\right)=0\, ,
\end{eqnarray}
where now we can see that the Ricci scalar introduces a a possibility for $k_\mu k^\mu=0$.

A remarkable feature of Eq.~\eqref{KG1} is that it is conformally invariant  \cite{wald} under the change in the metric $\tilde g_{\mu\nu}=\Omega^2 g_{\mu\nu}$ and in the field $\tilde\Phi=\Omega^{-1}\Phi$, for a general scalar function $\Omega=\Omega(x^\mu)$. An important consequence of this, is that Eq.~\eqref{KG1} can always be put in the flat spacetime form for conformally flat metric. As it was discussed in the Introduction, this means that for conformally flat spacetime, scalar waves described by Eq.~\eqref{KG1} always follows null geodesics. Notice that this is not true in the Klein--Gordon \eqref{KG0}, as that equation is not conformally invariant. This can be readily seen from Eqs.~\eqref{firstsetEqs2} for conformally flat metric $g_{\mu\nu}=\Omega^2\eta_{\mu\nu}$ (with $\eta_{\mu\nu}$ the flat spacetime metric), and $\phi=\Omega^{-1}\varphi$, with $\varphi$ constant. In this case, $R=-\eta^{\mu\nu}\partial_\mu\partial_\nu\Omega/\Omega^3=\Box\phi/\phi$. Thus, the scalar wave moves in null geodesics $k_\mu k^\mu=\Omega^{-2}\eta^{\mu\nu}k_\mu k_\nu=0$, with constant wavevector $k_\mu$. All these also solve identically the second equation \eqref{firstsetEqs2}.

An explicit example of this conformally invariant feature is the flat Friedmann--Robertson--Walker (FRW) cosmology in cartesian coordinates \cite{wald}. In this case the scalar of curvature is
\begin{equation}
R={6}\left(\frac{\ddot a}{a}+\frac{\dot a^2}{a^2}\right)\, ,
\end{equation}
where $a=a(t)$ the scale factor of the Universe \cite{wald}, and $\sqrt{-g}=a^3$.
In a generic  direction $\xi$, Eq.~\eqref{KG1} becomes
\begin{equation}
-\partial_0\left(a^3\dot\Phi\right)+a\partial_\xi^2\Phi-{a^3}\left(\frac{\ddot a}{a}+\frac{\dot a^2}{a^2}\right)\Phi=0\, ,
\end{equation}
which has a plane wave solution $\Phi=\phi \exp(iS)$, with amplitude and phase
\begin{equation}
\phi\propto\frac{1}{a}\, ,\qquad S=k\int\frac{dt}{a}\pm k\xi\, .
\end{equation}
The wavevectors $k_0=\partial_0S=k/a$ and $k_\xi=\partial_\xi S=k$, fulfill $k_\mu k^\mu=0$, and thus the plane wave follows null geodesics. Besides, the conservartion law, depicted in the second equation in \eqref{firstsetEqs2}, is identically satisfied.

Despite of the above, one must realize that if the metric is not conformally flat, then it is very difficult to show that the wave propagates in null geodesics \cite{Noonan}, eben in non-flat FRW cosmologies \cite{faraoni, faraoni2,zecca}. No general condition for Eq.~\eqref{KG1} has been found to establish such behavior. One of the simplest cases where a plane wave following null geodesics is not a solution of Eq.~\eqref{KG1} is for Bianchi I spacetime which describes an anisotropic cosmological Universe with metric $ds^2=-dt^2+a^2dx^2+b^2 dy^2+c^2 dz^2$ in cartesian coordinates \cite{ryan,Gron}, where $a$, $b$ and $c$ are arbitrary functions of time (the FRW limit is obtained when $a=b=c$). For the Bianchi I case, the conservation law in Eqs.~\eqref{firstsetEqs2} allow us to find a general solution for the time--dependent wave amplitude
\begin{equation}\label{solBiqnchi}
\phi=\phi_0 \left(a\, b\, c\,  k_0\right)^{-1/2}\, ,
\end{equation}
where $\phi_0$ is a constant. The amplitude is written in terms of the frequency that must be determined by the dispersion relation of Eqs.~\eqref{firstsetEqs2}. If we assume that the wave propagates along null geodesics $k_\mu k^\mu=0$, then the frequency has the form
\begin{equation}\label{assuNull}
k_0= \left(k_x^2/a^2+k_y^2/b^2+k_z^2/c^2\right)^{1/2}\, ,
\end{equation}
 where $k_x$, $k_y$ and $k_z$ can be choosen as constants. By using \eqref{assuNull} in Eq.~\eqref{solBiqnchi}, it is straightforward to show that $\Box\phi/\phi\neq R/6$ for Bianchi I spacetimes.
Therefore, the assumptions $k_\mu k^\mu=0$ and \eqref{assuNull} are incorrect, and the scalar wave does not propagate along null geodesics.

There are several other metrics where Eq.~\eqref{KG1} produces massless scalar waves  that do not propagate in null geodesics. One can wonder if other kind of coupling or modification of the massless Klein--Gordon equation can allow to define massless scalar waves in a way such that they propagates always in the null cone. In the next section we explore this possibility.

\section{New non-linear modified massless Klein--Gordon equation}
\label{secSup}

Eqs.~\eqref{KG0} and \eqref{KG1} have wave solutions that propagate inside the light cone.
In term of the wave vector $k_\mu$, the light cone propagation condition is $k^\mu k_\mu=0$, which may be written as
\begin{equation}\label{lightcone2}
g^{\mu\nu}\partial_\mu S\,  \partial_\nu S=0\, ,
\end{equation}
in terms of the phase of the wave. The construction of a new modified Klein--Gordon equation, which works both for flat and curved spacetimes, starts with the introduction of a Lagrangian density $\mathcal{L}=\mathcal{L}  \left(\rho(x^\alpha), S(x^\beta)\right)$ for a complex scalar field, in terms of the phase and  the density $\rho \equiv  \Phi \Phi^*$. This Lagrangian is
\begin{equation}\label{Lag2}
\mathcal{L}  =\frac{1}{2} \sqrt{-g}\,  \rho\, g^{\alpha\beta}\, \partial_\alpha S\, \partial_\beta S\, .
\end{equation}
Variations with respect to $\rho$ and $S$ gives rise to  the light cone propagation condition \eqref{lightcone2}, and to the conservation equation
\begin{equation}\label{conserv}
\partial_\alpha\left( \sqrt{-g} g^{\alpha\beta} \rho   \partial_\beta S\right)=0\, ,
\end{equation}
which is equivalent to the conservations laws in Eqs.~\eqref{firstsetEqs} and \eqref{firstsetEqs2}.
The Lagrangian \eqref{Lag2} produces the desired equations of motion: null geodesics propagation and the conservation law for the Klein--Gordon field. To put this in an explicit form, we can write the Lagrangian \eqref{Lag2} in terms of the original field variables $\Phi(x^\gamma)$ and its complex conjugate $\Phi(x^\gamma)^*$, to get
\begin{equation}\label{Lag3}
\mathcal{L} (\Phi,\Phi^*) = -\frac{1}{8} \sqrt{-g}  g^{\alpha\beta}\frac{(\Phi^*\Phi_{,\alpha}- \Phi{\Phi^*}_{,\alpha})(\Phi^*\Phi_{,\beta}- \Phi{\Phi^*}_{,\beta})}{\Phi\ \Phi^*} .
\end{equation}
Thus, variation of the Lagrangian density \eqref{Lag3} with respect to $\Phi^*$ yields
\begin{equation}\label{eqphiAH}
\Box\Phi-\frac{\Box \sqrt {\Phi \Phi^*}}{\sqrt {\Phi \Phi^*}} \Phi = 0\, ,
\end{equation}
whereas variations of \eqref{Lag3} with respect to $\Phi$ gives the complex conjugated of \eqref{eqphiAH}.

The modified Klein--Gordon equation \eqref{eqphiAH} is the most important result of this work. This equation is non--linear and fully general relativistic. It also has the remarkable property that it is conformally invariant in four--dimensional spacetimes under
the change $\tilde g_{\mu\nu}=\Omega^2 g_{\mu\nu}$ and $\tilde\Phi=\Omega^{-1}\Phi$, for an arbitrary function $\Omega(x^\mu)$. This can be explicitly seen by doing
\begin{eqnarray}
&&\tilde\Box\tilde\Phi-\frac{\tilde\Box \sqrt {\tilde\Phi \tilde\Phi^*}}{\sqrt {\tilde\Phi \tilde\Phi^*}} \tilde\Phi=\left(\frac{\Box\Phi}{\Omega^3}-\frac{\Phi\Box\Omega}{\Omega^4}\right)\nonumber\\
&&\qquad\qquad\qquad\qquad-\left(\frac{\Box\sqrt {\Phi \Phi^*}}{\Omega^3}-\frac{\sqrt {\Phi \Phi^*}\Box\Omega}{\Omega^4}\right)\frac{\Phi}{\sqrt {\Phi \Phi^*}}\nonumber\\
&&\qquad\qquad\qquad\quad=\frac{1}{\Omega^3}\left(\Box\Phi-\frac{\Box \sqrt {\Phi \Phi^*}}{\sqrt {\Phi \Phi^*}} \Phi\right)=0\, ,
\end{eqnarray}
by Eq.~\eqref{eqphiAH}.
Unlike Eq.~\eqref{KG0}, the modified massless Klein-Gordon equation \eqref{eqphiAH} always has scalar waves solutions travelling along null geodesics by construction. If the wave solution \eqref{complexdec} is used in \eqref{eqphiAH}, we straightforward obtain the equations
\begin{eqnarray}\label{firstsetEqs3}
k_\mu k^\mu=0\, ,\qquad
\nabla_\mu\left(\phi^2 k^\mu\right)=0\, .
\end{eqnarray}
Thereby, this new Klein--Gordon fields behaves as scalar photons. The results \eqref{firstsetEqs3} holds for any spacetime metric, and it is here where lies the importance of Eq.~\eqref{eqphiAH}. As far as we know, equations similar to \eqref{eqphiAH} have not been proposed before and we expect that they can bring new insights on the massless scalar field in curved spacetimes.

\section{Discussion}

The conformally coupled Klein--Gordon equation \eqref{KG1} reduces to the usual Klein--Gordon equation in flat spacetimes. On the contrary, the modified Klein--Gordon equation \eqref{eqphiAH} leads to a modification of the massless Klein--Gordon equation even in flat spacetime.
This behavior is due to the form of the new coupling term
\begin{equation}\label{GRBohm}
-\frac{\Box \sqrt {\Phi \Phi^*}}{\sqrt {\Phi \Phi^*}}\, ,
\end{equation}
which is not usual in General Relativity, but it often appears in quantum mechanics. The term \eqref{GRBohm} is  the general relativistic version of the well-known Bohm potential
\begin{equation}
V_{Bohm}=-\frac{\hbar^2}{2m}\frac{\nabla^2 \sqrt{n}}{\sqrt{n}}\, ,
\end{equation}
 that usually emerges in the Madelung-Bohm description of the non-relativistic Schr\"odinger equation \cite{madelung,bohm}, where $\hbar$ is the reduced Planck constant, $n$ is the probability density associated to the wavefunction,  $m$ is the mass of the field, and $\nabla^2$ is a Laplacian. The Bohm potential has been extensively explored in quantum mechanics \cite{holland, baggot, riggs, dennis}, solid state physics \cite{haasman,uzma,ferry}, and  quantum plasmas \cite{marklund, asenjoMahajan, taka,haas}.
Commonly, it is associated to the non--locality of quantum mechanics and it is a consequence of the wave--like nature of particles \cite{dennis}. The Bohm potential allows for the re--interpretation of quantum mechanics as the dynamics of a fluid with non--local interactions. This model has been laso studied even at relativistic quantum level \cite{asenjoFey} and explored in macroscopic--level experiments \cite{coud,harris,harris2,bush}.

 The meaning of the Bohm potential may be extrapolated to the Klein--Gordon equation as it models an extended object, a wave. In particular, as it was discussed in the Introduction, the EP does not guarantee that massless waves described by the Klein--Gordon equation on curved backgrounds follow null geodesics. A way to make sure that the null geodesic behavior always occurs for massless fields, is to force them to behave as  massless particles. In quantum mechanics, the wave characteristic of a quantum object is represented by the Bohm potential \cite{dennis}. Therefore, if a field is modelled by the modified Klein--Gordon equation which does not have the Bohm potential term, it will always behave as a particle. The term \eqref{GRBohm} removes the wave nature of the field, producing a a Klein--Gordon equation \eqref{eqphiAH} whose dynamics is equivalent to massless point particles on any curved (or flat) spacetime.

The interesting consequence of this interpretation of the coupling term \eqref{GRBohm} and Eq.~\eqref{eqphiAH}, is that suggest that any other form of coupling for the Klein--Gordon equation will always have some solutions that would have  propagation outside the light cones, i.e., tails.

It is interesting to remark the this new equation may be derivable from the current--current type Lagrangian \eqref{Lag3}. Eq.~\eqref{eqphiAH}
may provide new insights on the propagation of massless fields on any curved spacetime, such as inflationary ones. In addition, we believe that the modified Klein--Gordon equation can produce also interesting results for massive scalar fields. On the other hand, the identification of the Bohm potential as responsible for describing the wave--like features of any  field, hints that the same procedure may be applied to other wave equations, such as for spin fields. This will allow to describe massless fields propagating as point-like particles, where their spin feature is isolated from their wave nature.
Such studies will be pursued in the future.

\end{document}